\documentclass{article}
\pdfoutput=1
\usepackage{latexsym,amsmath,amssymb,lmodern,float,url}
\usepackage{jheppub}
\usepackage{natbib}
\usepackage{color}
\usepackage{multirow}
\usepackage{bm}
\usepackage{array}

\def\de{\delta^{\vphantom{1}}}
\def\bde{{\bar\delta}}
\def\qq{{q\bar q}}
\def\QQ{{Q\bar Q}}

\def\bt{{\bar\theta}}

\def\h3{{\displaystyle{\frac 3 2}}}
\def\tX{{\theta_{\! X}}}
\def\tZ{{\theta_{\! Z}}}

\begin{document}
\title{The Dynamical Diquark Model: Fine Structure and Isospin}
\author{Jesse F. Giron}
\emailAdd{jfgiron@asu.edu}
\author{Richard F. Lebed}
\emailAdd{Richard.Lebed@asu.edu}
\author{Curtis T. Peterson}
\emailAdd{curtistaylor.peterson@asu.edu}
\affiliation{Department of Physics, Arizona State University, Tempe,
AZ 85287, USA}

\date{July, 2019}

%%%%%%%%%%%%%%%%%%%%%%%%%%%%%%%%%%%%%%%%%%%%%%%%%%%%%%%
\abstract{
We incorporate fine-structure corrections into the dynamical diquark
model of multiquark exotic hadrons.  These improvements include
effects due to finite diquark size, spin-spin couplings within the
diquarks, and most significantly, isospin-dependent couplings in the
form of pionlike exchanges assumed to occur between the light quarks
within the diquarks.  Using a simplified two-parameter interaction
Hamiltonian, we obtain fits in which the isoscalar $J^{PC} \! =
\! 1^{++}$ state---identified as the $X(3872)$---appears naturally as
the lightest exotic (including all states that are predicted by the
model but have not yet been observed), while the closed-charm decays
of $Z_c(3900)$ and $Z_c(4020)$ prefer $J/\psi$ and $h_c$ modes,
respectively, in accord with experiment.  We explore implications of
this model for the excited tetraquark multiplets and the pentaquarks.}
%%%%%%%%%%%%%%%%%%%%%%%%%%%%%%%%%%%%%%%%%%%%%%%%%

\keywords{Exotic hadrons, diquarks, lattice QCD}
\maketitle

\section{Introduction}
\label{sec:Intro}

The census of heavy-quark ($c$- or $b$-containing) exotic hadrons has
now reached about 40 candidates, with no indication of a slackening in
the pace of their discovery.  Equally surprising is that no single
theoretical picture has emerged as a global paradigm to describe their
structure.  Advocates can point to examples among the exotics for
which hadronic molecules, hybrids, hadroquarkonium, diquark states, or
threshold effects are particularly well suited, while detractors can
point to equally compelling counterexamples.  The status of both
experimental results and theoretical pictures have been reviewed
extensively in a number of recent
reviews~\cite{Lebed:2016hpi,Chen:2016qju,Hosaka:2016pey,
Esposito:2016noz,Guo:2017jvc,Ali:2017jda,Olsen:2017bmm,
Karliner:2017qhf,Yuan:2018inv,Liu:2019zoy,Brambilla:2019esw}.

From the theoretical point of view, all of the pictures are based upon
sound ideas---phenomena either proven to exist in phenomenology ({\it
e.g.}, atomic nuclei as hadronic molecules) or as well-studied
features of quantum field theory [{\it e.g.}, the SU(3)$_{\rm
color}$-triplet diquark attraction; hadronic on-shell
threshold-induced singularities in Green's functions from chiral
Lagrangians].  However, which particular mechanisms are most important
to describe the detailed spectrum and decay modes of the existing
exotics remains an unsettled and hotly disputed question.  Even if one
specific picture eventually emerges as the dominant model, very likely
the inclusion of modifications due to the other effects---{\it i.e.},
full coupled-channel calculations---will be essential in order to
obtain a detailed understanding of the exotics.

In this spirit, it is essential to push any given theoretical picture
to its limit, examining both its successes and shortcomings as a
global model for the exotics.  The purpose of the current work is to
continue the development of the {\it dynamical diquark picture\/} of
exotics~\cite{Brodsky:2014xia,Lebed:2015tna}, which is defined through
the color attraction of the channel ${\bm 3} \otimes {\bm 3} \to
\bar{\bm 3}$ to form heavy-light diquarks $\de \! \equiv \!
(Q q)_{\bar{\bm 3}}$ and their antiparticles $\bde \! \equiv \! (\bar
Q \bar q)_{\bm 3}$ as quasi-bound hadronic subcomponents.  The
presence of a heavy quark $Q$ means that $\de$ is more spatially
compact than a typical light-quark hadron, while the large energy
release available in the production processes of exotics (either
through $b$-quark decay [for charmoniumlike states] or collider
production) means that the color-nonsinglet $\de$ and $\bde$ can
separate a sufficient distance to allow the $\de$-$\bde$ state to
temporarily evade color recombination (into, {\it e.g.}, a hadronic
molecule) until the quarks of $\de$ ultimately combine with the
antiquarks of $\bde$ in the decay of the state.  In other words,
Ref.~\cite{Brodsky:2014xia} argues that when a collection of quarks
$\QQ \qq$ forms, the dominant decay mode is indeed two mesons, with
the formation of a di-meson hadronic molecule being a rare special
case.  But if $Q , q$ initially lie closer to each other than to $\bar
Q , \bar q$, then $\de$ and $\bde$ formation is an alternative.  If
the $\de$-$\bde$ pair were created with low relative momentum within a
typical hadronic volume, then such a state would still exhibit a large
overlap with the two-meson wave function.  In that case nothing would
impede a very rapid decay to this mode, obscuring the fact that a
$\de$-$\bde$ pair had initially formed.  But if a large relative
momentum occurs in the production process to push the diquarks apart
before this recombination can occur, then the two-meson wave-function
overlap is suppressed, the decay is delayed, and the appearance of a
$\de$-$\bde$ state becomes perceptible.  Put another way, the
momentum-space wave function of the two diquarks overlaps more
strongly with two-meson states in its small-momentum regime, and with
$\de$-$\bde$ states in its large-momentum regime.  Moreover, the
triplet-channel attraction need not conclude after just two
quarks~\cite{Brodsky:2015wza}, leading to the proposal of {\it
triquarks} $\bt \! \equiv \! \left[ \bar Q (q_1 q_2)_{\bar {\bm 3}}
\right]_{\bm 3}$ as components of pentaquark states in the combination
$\bt \de$~\cite{Lebed:2015tna}.

The dynamical diquark picture has been developed into a full model,
including a specific spectroscopy and decay selection rules, in
Ref.~\cite{Lebed:2017min}.  The key ingredient necessary to
characterize states formed from separated $\de$-$\bde$ or $\bt$-$\de$
pairs is the introduction of the Born-Oppenheimer (BO)
approximation~\cite{Born:1927boa}, which distinguishes the heavy,
slowly changing $\de$ and $\bde$ (or $\bt$) from the rapidly changing
degrees of freedom in the color flux tube connecting them.  The
spectrum of flux-tube configurations of nontrivial gluon content has,
in turn, been studied on the lattice for decades; for example, these
simulations have been used to compute heavy-quarkonium hybrid-meson
masses~\cite{Liu:2012ze}.

In Ref.~\cite{Giron:2019bcs}, the results of lattice simulations
obtained by two independent
collaborations~\cite{Juge:2002br,Capitani:2018rox} for two separated,
color-triplet sources have been input as static-source BO potentials
$V(r)$ for Schr\"{o}dinger equations of $\de$-$\bde$ and $\bt$-$\de$
systems.  Any observed exotic of known mass and $J^{PC}$ quantum
numbers may then be identified with a state of the same $J^{PC}$
appearing in one of the multiplets listed in Ref.~\cite{Lebed:2017min}
and selected as a reference state, its mass serving as a particular
eigenvalue of the Schr\"{o}dinger equations, which for tetraquarks
fixes the diquark mass $m_\de$.  But then, with $V(r)$ and $m_\de$
specified, the {\em entire\/} mass spectrum of all tetraquarks is
completely determined---at least, ignoring the fine-structure mass
splittings within the levels of each BO potential.  If one chooses the
$1^{++}$ $X(3872)$ to fix the (positive-parity) ground-state multiplet
$\Sigma^+_g(1S)$, then Ref.~\cite{Giron:2019bcs} shows that the
(negative-parity) first excited levels $\Sigma^+_g(1P)$ appear at
about 4240~MeV, in excellent agreement with $1^{--}$ states such as
the $Y(4220)$ appearing nearby, and the next (positive-parity) excited
levels $\Sigma^+_g(2S)$ appear at about 4440~MeV, in excellent
agreement with the appearance of the $1^{+-}$ state $Z_c(4430)$.
Pentaquarks can then be studied by using the value of $m_\de$ obtained
from the tetraquark fit to select a reference pentaquark state to fix
$m_{\bt}$, and hence, predict the rest of the spectrum.

To go further with this analysis, however, one must consider the
aforementioned fine-structure corrections.  Just as for quarkonium,
one can identify multiple types of such corrections: spin-spin,
spin-orbit, tensor, Darwin terms, {\it etc.}.  However, multiquark
exotics offer a much richer possible set of interactions, simply due
to the greater combinatorics available to their constituent particles.
Choosing to work with a diquark model simplifies matters somewhat, by
clustering the components into identifiable subunits with good quantum
numbers.  For example, Ref.~\cite{Maiani:2014aja} achieved rather
satisfying results in their ``Type-II'' diquark model by assuming that
the dominant spin-spin interactions are solely those between the
quarks within each diquark; the mass splitting between the two
$1^{+-}$ states $Z_c^0(3900)$ and $Z_c^0(4020)$ arises quite naturally
in this scheme.

One ingredient that, to our knowledge, has not before been included in
previous diquark models is isospin dependence in the interaction
potential.  In the most naive type of tetraquark model, two quarks and
two antiquarks are placed in close proximity, and (in the limit $m_u
\! = \! m_d$) one expects no distinction between tetraquarks differing
only in the light-flavor contents $u\bar u$, $u \bar d$, $d \bar u$,
and $d \bar d$.  That is, one expects completely degenerate quartets
consisting of $I \! = \! 0$ and $I \! = \! 1$ multiplets.  But the
physical exotics appear to form ordinary $I \! = \! 0$ and $I \! = \!
1$ multiplets (the experimental absence~\cite{Choi:2011fc} of a
charged partner to the $X(3872)$ is particularly significant in this
respect), so a truly predictive model of exotics {\em must\/} contain
isospin-dependent effects at some level.  Since the $\de$-$\bde$ pair
is connected strongly by color-nonsinglet interactions, one expects
the same for the isospin-exchange quanta in this model (Such an
assumption however is not strictly necessary for the analysis
presented below; the phenomenological consequences of assuming purely
color-singlet isospin exchanges between the component quarks of
hadrons have been successfully studied for quite some
time~\cite{Glozman:1995fu}).  In the context of dense QCD, a variant
of the Nambu-Goldstone theorem has been
demonstrated~\cite{Alford:1998mk}, which means that light pionlike
exchange (colored, in this case) can exist in settings other than that
of the zero-density environment between color-singlet hadrons.  If one
posits that the interior of the color flux tube connecting the
$\de$-$\bde$ pair is another such environment, then light ``pions'',
possibly partly colored, could exist and propagate across the color
flux tube, evading the strong color screening that might impede the
propagation of ordinary color-singlet mesons, and providing the
essential isospin dependence in these states.  We emphasize that such
an effect is purely conjectural at this stage, but discuss later how
its existence might be established on the lattice.

In the ground-state multiplet $\Sigma^+_g(1S)$, the 6 possible states
[see Eq.~(\ref{eq:Swavediquark}) or (\ref{eq:HQbasis}) below] should
therefore actually be listed as 6 isosinglets and 6 isotriplets, for a
total of 12 mass eigenstates (when $m_u \! = \!  m_d$).  Likewise, one
finds 28 mass eigenstates for the first excited [$\Sigma^+_g(1P)$]
multiplet and another 12 for the second excited [$\Sigma^+_g(2S)$]
multiplet.  Such large multiplicities have led to the most frequent
criticism of diquark models, that they tend to overproduce states
compared to experiment.

In this regard, however, several points should be noted: First, new
exotic states are still being discovered or resolved---even at
relatively low masses---virtually every year, so it is not at all
impossible that the final tally in any flavor sector may turn out to
be well over 100\@.  Second, some of the predicted states have
$J^{PC}$ quantum numbers that may be difficult to probe with available
production channels ({\it e.g.}, the conventional $\psi_3(1D)$
($3^{--}$) charmonium candidate state $X(3842)$ has only been observed
for the first time this year~\cite{Aaij:2019evc}).  Third, if a state
lies only a modest amount above its fall-apart decay threshold, then
it can be quite wide, and possibly difficult to distinguish from
background ({\it e.g.}, the conventional charmonium $\chi_{c0}(2P)$
candidate state $\chi_{c0}(3860)$ lies only about 130~MeV above the
$D\bar D$ threshold but has a width of about
200~MeV~\cite{Chilikin:2017evr}, which made it challenging to resolve
until relatively recently).

The second common criticism of such models is that the diquark
quasiparticles are not pointlike (estimated radii of a few times
0.1~fm~\cite{Brodsky:2014xia}), and if the full exotic states are not
too many times larger, then the $\de$ and $\bde$ wave functions must
have considerable spatial overlap.  But then, one expects that the
stronger $\qq$ color-singlet attractions should lead to a
rearrangement of the quark constituents into a configuration
resembling a hadron molecule or hadroquarkonium (see, {\it
e.g.},~\cite{Voloshin:2019rfo}).  That is to say, for small
$\de$-$\bde$ separations (or small heavy-meson separations in a
molecular model), the $\QQ$ pair lie in close proximity within the
cloud of the $\qq$ pair, and the nature of the wave function resembles
that of hadroquarkonium, with these three pictures distinguished only
by the specific color correlations of the quark pairs. In the original
dynamical diquark model, such a color reorganization prior to decay is
suppressed by the separation of the $\de \bde$ pair.  One can also
develop models in which this separation is not merely the result of
the production process, but is enforced by a potential
barrier~\cite{Maiani:2019cwl}.

In this work we also explore the effect of finite diquark sizes by
modeling the Schr\"{o}dinger equations to transition at a chosen
distance $R$ from ones describing the interaction of the $\de \bde$
pair to ones describing just the interaction of the $\QQ$ pair.  Then
the exotic consists primarily of an interacting $\QQ$ pair residing in
a shell of constant potential provided by the light $\qq$ pair and
glue, which is indeed quite similar to the hadroquarkonium picture.
We see below that the calculated spectrum is fairly insensitive to
changes of $R$ from zero to physically reasonable values, providing
confidence in this aspect of the modeling of $\de$-$\bde$ states.

This paper is organized as follows.  In Sec.~\ref{sec:Finite} we
examine the effect of finite diquark size on the exotics spectrum in
the manner just described.  The introduction of isospin-dependent
interactions between the $\de$-$\bde$ pair appears in
Sec.~\ref{sec:Isospin}, and we compute the corresponding expressions
for the spectrum of the ground-state $\Sigma^+_g(1S)$ multiplet,
including both isospin and spin-spin dependence.  In
Sec.~\ref{sec:Results} we fit the $X(3872)$, $Z_c(3900)$, and
$Z_c(4020)$ states to the model parameters, and show that natural
choices of the unfixed parameters allow all unconfirmed members of the
multiplet to lie higher in mass, and indeed respect the pattern of
$Z_c(3900)$/$Z_c(4020)$ closed-charm decay modes.  Finally, in
Sec.~\ref{sec:Discussion} we indicate the direction of the analogous
investigation for excited multiplets, pentaquarks, and the $b\bar b$
sector, and summarize our findings.

\section{Effects Due to Finite Diquark Size}
\label{sec:Finite}

The calculations of Ref.~\cite{Giron:2019bcs} assume a potential
$V(r)$ valid for a $\de$-$\bde$ pair that can assume any separation
$r$.  The functional form $V(r)$ is taken from lattice simulations for
a heavy (hence static) particle pair transforming as ${\bm 3}$ and
$\bar{\bm 3}$ under SU(3)$_{\rm color}$.  The specific masses, spin
statistics, flavor, and charge quantum numbers of the heavy sources
are considered immaterial to the results of these calculations, and so
one may use the same potentials for heavy $\QQ$ states (using the
ground-state BO potential $\Sigma^+_g$ of the color flux tube) or
their hybrids $\QQ g$ (using the excited BO potentials such as
$\Pi^+_u$, $\Sigma^-_u$, {\it etc.}), or for $\de \bde$ tetraquark and
$\bt \de$ pentaquark states.  Of course, quarks are fundamental,
presumably pointlike constituents, while diquarks and triquarks have a
finite spatial extent.  One should not expect that the same potential
$V(r)$ as used for interactions between pointlike sources should hold
$\de$-$\bde$ or $\bt$-$\de$ pairs at arbitrarily small values of $r$,
in regions where the wave functions of the quasiparticles strongly
overlap.

We present a simple proposal to test the effect of the finite diquark
(or triquark) size: Since each such quasiparticle in this model
contains exactly one heavy quark or antiquark, we suppose for
simplicity the existence of a critical separation $R$ between the
centers of the $\de$-$\bde$ or $\bt$-$\de$ pair, at which point the
wave-function overlap between the two is considered significant.  Were
the quasiparticles hard spheres, then $R$ would equal the sum of their
radii.  At distances $r \! < \! R$, we suppose that the dominant
interaction becomes the attraction between the $\QQ$ pair, which uses
precisely the same $V(r)$ as for $\de \bde$ or $\bt \de$ since it is
also a ${\bm 3}$-$\bar{\bm 3}$ pair.  However, the masses appearing in
the kinetic-energy term of the Schr\"{o}dinger equation are no longer
$m_\de$ or $m_\bt$, but $m_Q$.  We further suppose that, at reasonably
small $R$, the $\qq$ pair simply provides a constant potential in
which the $\QQ$ pair interact.  Since the state then consists of a
$\QQ$ pair within a light cloud consisting of the $\qq$ pair and glue,
the physical picture becomes quite similar to that of
hadroquarkonium~\cite{Dubynskiy:2008mq}.

Explicitly, the Hamiltonian used in the Schr\"{o}dinger equation
assumes the usual form $H = \frac{p^2}{2\mu} + V(r)$, where $p$ is the
relative momentum of the constituents, and $V(r)$ is the specific
lattice-computed ${\bm 3}$-$\bar{\bm 3}$ potential chosen for the
calculation holding for all $r$.  However,
\begin{equation}
\frac{1}{\mu} = \left\{
\frac{1}{m_\de} + \frac{1}{m_\bde} \ {\rm for} \  r  > R \, , \
{\rm and} \
\frac{1}{m_Q} + \frac{1}{m_{\bar Q}} \ {\rm for} \ r < R \right\} \, ,
\end{equation}
and matching at $r \! = \! R$ is accomplished by imposing continuity
of the eigenfunction and its first derivative.  The values of
$m_{\de}$, $m_{\bde}$ are then adjusted to obtain the physical mass
eigenvalue [{\it e.g.}, $m_{X(3872)}$].  One may of course introduce
any one of a number of different methods with a variety of refinements
to incorporate the finite size of the diquark, but this simple ansatz
provides a convenient one-parameter ($R$) method of testing the
limitations of the approach.

At $R \! = \! 0$, the diquark becomes pointlike.  One then recovers
the results calculated in Ref.~\cite{Giron:2019bcs}, specifically the
first fits of Table~3 (within small numerical tolerances), in which
the $\Sigma^+_g(1S)$ mass eigenvalue is fixed to that of the
$X(3872)$, the diquark mass $m_\de$ entering the Schr\"{o}dinger
equation is obtained as an output, and the charm-quark mass is fixed
to a typical value, $m_c \! = \! 1.477$~GeV~\cite{Berwein:2015vca}.
We have computed modifications to the spectrum using the above ansatz
and a variety of values of $R$ ranging from $0 \! \to \! 1$~fm
(corresponding to a classical hard-sphere diquark radius of 0.5~fm).
Sample results are presented in Table~\ref{tab:FiniteDiquark}; the
right-hand columns ($R \! = \!  0.0$~fm) reproduce the results of
Ref.~\cite{Giron:2019bcs}, and the left-hand columns are computed at
$R \! = \! 0.7$~fm\@.  The acronyms refer to the results of lattice
simulations by two collaborations,
JKM~\cite{Juge:2002br,Morningstar:2019} and
CPRRW~\cite{Capitani:2018rox}.
\begin{table}[ht]
\caption{Mass eigenvalues $M$ (in GeV) for hidden-charm
dynamical diquark states that are eigenstates (with quantum numbers
$nL$) of a Schr\"{o}dinger equation in which $V(r)$ is the
ground-state BO potential $\Sigma^+_g$.  The functional form of $V(r)$
is given by lattice simulations
JKM~\cite{Juge:2002br,Morningstar:2019} or
CPRRW~\cite{Capitani:2018rox}.  The eigenvalue for the $1S$ state is
fixed to the $X(3872)$ mass, and the diquark mass $m_{\delta}$ (in
GeV) is the parameter that must be used as input for the
Schr\"{o}dinger equation in order to achieve this constraint.  As
described in the text, this equation uses $m_{\delta}$ as its mass
parameter for $r \! > \! R$, and $m_c \! = \! 1.477$~GeV for $r \! <
\! R$ (the separation between $\delta$ and $\bar \delta$ centers).
Also computed are the corresponding expectation values for the length
scales ${\langle 1/r \rangle}^{-1}$ and $\langle r \rangle$ (in fm).}
\label{tab:FiniteDiquark}
{\vskip 0.5em}
\setlength{\extrarowheight}{1.3ex}
\begin{tabular}{c @{\hskip 0.5em} c | c c c @{\hskip 1.0em} c
@{\hskip 0.5em} | c c c @{\hskip 1.0em} c @{\hskip 0.5em}}
%\begin{tabular}{c c | c c c c | c c c c }
\hline\hline
\multicolumn{2}{c|}{\multirow{2}{*}{}}&\multicolumn{4}{c|}{$R=0.7$ fm}
& \multicolumn{4}{c}{$R=0.0$ fm}\\\cline{3-4}
\hline
BO states & Potential & $M$ & $m_{\delta}$&
${\langle}{1/r}{\rangle}^{-1}$ & ${\langle}r{\rangle}$ & $M$ &
$m_{\delta}$ &
${\langle}{1/r}{\rangle}^{-1}$ & ${\langle}r{\rangle}$ \\
\hline
%Beginning \Sigma_g^+(1S)
\multicolumn{1}{ c }{{$\Sigma_g^+(1S)$}}
&JKM& $\textbf{3.8716}$ & $1.8556$ & $0.36925$ & $0.32136$ &
$\textbf{3.8716}$ & $1.8750$ & $0.27202$ & $0.36461$ \\
&CPRRW& $\textbf{3.8717}$ & $1.8390$ & $0.36780$ & $0.32538$ &
$\textbf{3.8716}$ & $1.8532$ & $0.27521$ & $0.36915$ \\
%Beginning \Sigma_g^+(2S)
\multicolumn{1}{ c }{{$\Sigma_g^+(2S)$}}
&JKM& $4.4231$ & $1.8556$ & $0.49495$ & $0.65605$ & $4.4435$ &
$1.8750$ & $0.42698$ & $0.69081$ \\
&CPRRW& $4.4256$ & $1.8390$ & $0.49385$ & $0.66085$ & $4.4405$ &
$1.8532$ & $0.43064$ & $0.69640$ \\
%Beginning \Sigma_g^+(1P)
\multicolumn{1}{ c }{{$\Sigma_g^+(1P)$}}
&JKM& $4.2067$ & $1.8556$ & $0.60909$ & $0.50400$ & $4.2462$ &
$1.8750$ & $0.48962$ & $0.56613$ \\
&CPRRW& $4.2072$ & $1.8390$ & $0.60589$ & $0.50798$ & $4.2429$ &
$1.8532$ & $0.49376$ & $0.57067$ \\
%Beginning \Sigma_g^+(1D)
\multicolumn{1}{ c }{{$\Sigma_g^+(1D)$}}
&JKM& $4.4863$ & $1.8556$ & $0.75652$ & $0.67579$ & $4.5323$ &
$1.8750$ & $0.66419$ & $0.73132$ \\
&CPRRW& $4.4881$ & $1.8390$ & $0.75436$ & $0.67993$ & $4.5277$ &
$1.8532$ & $0.66931$ & $0.73656$ \\
\hline
\end{tabular}
\end{table}

One immediately notes how little many of the numerical results change.
The value of $m_\de$, for example, decreases by a percent or less.
The $2S$-$1S$ mass splitting decreases by only 15--20~MeV in going
from $R \! = \! 0.0$~fm to $R \! = \! 0.7$~fm, the $1P$-$1S$ splitting
decreases by $36$--$40$~MeV, and even the $1D$-$1S$ splitting
decreases by no more than 46~MeV\@.  These changes amount to roughly
3--12\% decreases in the overall size of the splittings, with the
largest effect occurring in the $1P$-$1S$ splitting.  It is only for
$R \! > \!  0.8$~fm that one begins to see the results changing more
dramatically, so we take 0.4~fm as an indication of the largest
diquark radius one may reasonably use as quasiparticles in these
calculations.  The length-scale expectation values, on the other hand,
change quite drastically with $R$; but since $\left< r \right>$, for
example, is a convolution of the average distance between the
$\de$-$\bde$ pair (for $r \! > \! R$) with the average distance
between the $\QQ$ pair (for $r \! < \! R$), it is not surprising that
$\left< r \right>$ is sensitive to changing the mass parameter in the
Schr\"{o}dinger equation from $m_\de$ to $m_Q$.

\section{Isospin Interactions Between Diquarks}
\label{sec:Isospin}

The one-pion exchange potential between two spin-$\frac 1 2$ nucleons
(with corresponding spin ${\bm \sigma}$ and isospin ${\bm \tau}$
operators), separated by a relative position vector ${\bm r}$, has
been known for many decades (arguably, as early as
1938~\cite{Kemmer:1938a,Kemmer:1938b}).  In modern notation, it reads:
%
%\begin{widetext}
\begin{equation}
\label{eq:OnePion}
V_{\pi} ( {\bm r} ) = \left( \! \frac{g_A}{\sqrt{2}f_\pi} \! \right)^2
\! {\bm \tau}_1 \cdot {\bm \tau}_2 \left[ \frac{m_\pi^2}{12\pi}
\frac{e^{- m_\pi r}}{r} \left(  {\bm \sigma_1} \! \cdot {\bm \sigma}_2
+ S_{12} \! \left[ 1 + \frac{3}{m_\pi r} + \frac{3}{(m_\pi r
)^2} \right] \right) - \frac{1}{3} \, {\bm \sigma}_1 \! \cdot
{\bm \sigma}_2 \, \delta^{(3)} ( {\bm r} )
\right]\,,
\end{equation}
%\end{widetext}
%
where the tensor operator $S_{12}$ is defined by
\begin{equation}
\label{eq:Tensor}
S_{12} \equiv 3 \, {\bm \sigma}_1 \! \cdot {\bm r} \, {\bm \sigma}_2
\! \cdot {\bm r} / r^2 - {\bm \sigma}_1 \! \cdot {\bm \sigma}_2 \, .
\end{equation}
In particular, each term depends upon isospin exchange (${\bm \tau}_1
\! \cdot \! {\bm \tau}_2$), as well as upon spin exchange,
between the nucleons.  $S_{12}$ is a rank-2 tensor operator in both
spin and position space, and therefore by the Wigner-Eckart theorem
all states in an $S$ wave connect through $S_{12}$ only to $D$-wave
states, which are expected to lie much higher in
energy~\cite{Giron:2019bcs}.  $P$-wave states, on the other hand, have
nonvanishing diagonal $S_{12}$ matrix elements.  The contact term
$\delta^{(3)} ( {\bm r} )$ is included for formal reasons, but the
one-pion exchange potential has long been known~\cite{Taketani:1951}
to require major modifications at separations below about 2~fm, so
that the $\delta^{(3)} ( {\bm r} )$ term should actually be replaced
by explicit short-distance effects.  In any case, this term still
carries the same spin and isospin dependence as the long-distance
potential.

Since the $\sim \! 2$~fm range approximately equals the sum of two
nucleon radii (as indicated by, {\it e.g.}, their $\approx \!
0.86$~fm magnetic radii~\cite{Tanabashi:2018oca}), one may suppose
that a major factor in the transition from the one-pion-exchange
region to that of heavier-meson or multi-pion exchanges is the
appearance of a substantial overlap of nucleon wave functions.

In Eq.~(\ref{eq:OnePion}), the experimental value of the axial
nucleon-pion coupling is $g_A \!  = \! 1.2732(23)$, and the pion decay
constant (in this normalization) is $f_\pi \! = \!
130.2(1.7)$~MeV~\cite{Tanabashi:2018oca}.  Using an isospin-averaged
pion mass, we find
\begin{equation} \label{eq:PionCoeff}
\left( \! \frac{g_A}{\sqrt{2}f_\pi} \! \right)^2 \!
\frac{m_\pi^2}{12\pi} = 4.72 \; {\rm MeV} \cdot {\rm fm} \, ,
\end{equation}
a value to be used below as a comparison with the strength of the
isospin exchange between diquarks.

The exchange of pions (and other mesons) between color-singlet hadrons
to bind hadronic mole\-cules, both in the form of potential exchanges
such in as the $NN\pi$ interaction discussed above, and in
calculations employing chiral Lagrangians, has long been one of the
primary mechanisms used to study multiquark exotic
hadrons~\cite{Guo:2017jvc}.  The long range of pion interactions of
course stems from its status as the lightest meson, which in turn
follows from its role as a Nambu-Goldstone boson of chiral symmetry
breaking.  The long-distance isospin dependence of the interactions in
molecular models follows primarily from the isospin content of
exchanged pions.

We now turn to the analogous interaction for diquarks.  In the
eigenstates of the dynamical diquark model, the $\de$-$\bde$ pair
assume a nonzero separation, and each of $\de$, $\bde$ contains a
light quark that carries isospin $I \! = \! \frac 1 2$.  However,
$\de$ and $\bde$ are color nonsinglets, and they are connected by a
color flux tube.  The question then becomes whether isospin-dependent
exchanges can occur in the environment of nonzero color charge.  In
fact, a related question was addressed some time ago in the context of
high-density QCD\@.  As shown in the context of {\em color-flavor
locking}, the Nambu-Goldstone theorem of chiral-symmetry breaking
remains valid even within this environment of high quark density, so
that colored analogues of pions have been shown to exist in this
case~\cite{Alford:1998mk}.  In this work we propose that a similar
effect arises along an extended color flux tube: In this dense,
colored environment of extended spatial size, gluons remain the
dominant component, but the formation of a partly colored quark
condensate subcomponent in this non-vacuum environment is a very real
physical possibility.  An analogue to pion exchange would then exist
between the $\de$-$\bde$ pair, providing a natural source of isospin
dependence in the exotics spectrum of the dynamical diquark model.  To
be clear, we do not take this effect in the current scenario of the
$\de$-$\bde$ interaction in any sense to be proven to exist, but we do
consider such an isospin-dependent interaction with pionlike couplings
to be a plausible physical phenomenon.  Lattice simulations could
provide evidence for such an effect: One could, for example, fix in
space two heavy diquarks carrying nontrivial isospin quantum numbers,
calculate their interaction potential energy, and examine whether any
part of it is flavor non-universal.

Assuming then that the color flux tube connecting the $\de$-$\bde$
pair supports exchanges of a Nambu-Goldstone boson, one expects a
potential interaction between the light flavors within the diquarks
similar in form to Eq.~(\ref{eq:OnePion}).  In this paper we ignore
the $S_{12}$ term, since the fit is confined to the lowest $S$-wave
multiplet.  However, nothing in principle prevents an analysis of the
$P$-wave (or higher) states; only a paucity of confirmed states in
this multiplet discourages such a study at this time, and we provide a
few relevant comments on a $P$- or higher-wave analysis in
Sec.~\ref{sec:Discussion}.  Neglecting $S_{12}$, the remaining terms
of Eq.~(\ref{eq:OnePion}) are proportional to ${\bm
\tau}_q \! \cdot \! {\bm \tau}_{\bar q} \; {\bm \sigma}_q \! \cdot \!
{\bm \sigma}_{\bar q}$.  In the current model, we simply label the
coefficient of this operator as $V_0$.

One may take the phenomenology a step further in order to compare to
ordinary one-pion exchange.  As discussed above, we strike the contact
term $\delta^{(3)} ( {\bm r} )$ from the exchange potential, since at
short distances the diquark wave functions must overlap, necessarily
leading to a more complicated interaction.  Next, since the
calculations of Ref.~\cite{Giron:2019bcs} show how to compute any
expectation value without the need of calculating explicit wave
functions, one may obtain an explicit expectation value for the Yukawa
part of the potential, $\langle e^{-m_\pi r} / r \rangle$, and from
this result, extract a coefficient called $\tilde V_0$ that may be
compared with the combination in Eq.~(\ref{eq:PionCoeff}).
Explicitly, we write the full isospin-dependent potential $V_I(r)$ as
\begin{eqnarray}
V_I (r) & = & \tilde{V}_0 \, \times \frac{e^{-m_\pi r}}{r} \times
{\bm \tau}_q \! \cdot \! {\bm \tau}_{\bar q} \;
{\bm \sigma}_q \! \cdot \! {\bm \sigma}_{\bar q}  \, , \nonumber \\
V_0 & \equiv & \tilde{V}_0 \, \left< \frac{e^{-m_\pi r}}{r} \right> \,
.
\label{eq:V0tilde}
\end{eqnarray}
It bears mentioning that the Yukawa potential expectation value
decreases for excited states, and so while one may suppose that
${\tilde V}_0$ should be approximately the same constant for all
multiplets, the particular value of $V_0$ obtained below
[Eq.~(\ref{eq:FinalParams})] holds only for the $\Sigma^+_g(1S)$
multiplet.  An analogous effect explains why fine-structure splittings
in ordinary quarkonium decrease for higher multiplets.  A direct
comparison with Eq.~(\ref{eq:PionCoeff}) is also difficult because the
relation between the observed (vacuum) pion mass $m_\pi$ and the mass
parameter for the corresponding in-medium exchange quantum (the
``partly colored pion'') along the color flux tube is unknown, not to
mention the size of its coupling to the diquark (the analogue to
$g_A/f_\pi$).  Purely for sake of comparison, we take the mass of the
``partly colored pion'' in Eq.~(\ref{eq:V0tilde}) to equal $m_\pi$,
even the true mechanism of isospin exchange might be quite different.
One may expect on the basis of ordinary hadronic phenomenology that
the proper mass scale of the colored exchange is rather larger (at
least several hundred MeV), but using such a value does not radically
alter our numerical results.

%For
%example, even though a partly colored pion is forbidden by confinement
%from appearing in the vacuum, it could appear in some environments
%that are not especially dense with color charge, such as the overlap
%regions between nucleons in a complex nucleus. Such an effect could
%lead to noticeable modifications of known phenomenology.

The states in the ground-state multiplet $\Sigma^+_g(1S)$, prior to
introducing isospin, are defined as~\cite{Maiani:2014aja}
\begin{eqnarray}
J^{PC} = 0^{++}: & \; & X_0 \equiv \left| 0_\de , 0_\bde \right>_0 \,
, \ \
X_0^\prime \equiv \left| 1_\de , 1_\bde \right>_0 \, , \nonumber \\
J^{PC} = 1^{++}: & & X_1 \equiv \frac{1}{\sqrt 2} \left( \left| 1_\de
, 0_\bde \right>_1 \! + \left| 0_\de , 1_\bde \right>_1 \right) \, ,
\nonumber \\
J^{PC} = 1^{+-}: & & \  \, Z \equiv \frac{1}{\sqrt 2} \left( \left|
1_\de , 0_\bde \right>_1 \! - \left| 0_\de , 1_\bde \right>_1 \right)
\, , \nonumber \\
& & \; Z^\prime \equiv \left| 1_\de , 1_\bde \right>_1 \, ,
\nonumber \\
J^{PC} = 2^{++}: & & X_2 \equiv \left| 1_\de , 1_\bde \right>_2 \, ,
\label{eq:Swavediquark}
\end{eqnarray}
where the number preceding each $\de$($\bde$) subscript is the diquark
(antidiquark) spin ($s_\de$ and $s_{\bde}$, respectively), while the
outer subscript on each ket is the total quark spin $J$.  In terms of
the basis of good $\qq$ and $\QQ$ spin quantum numbers ($s_{\qq}$ and
$s_{\QQ}$, respectively), the corresponding eigenstates are
\begin{eqnarray}
{\tilde X}_0 & \equiv & \left| 0_\qq , 0_\QQ \right>_0 =
+ \frac{1}{2} X_0 + \frac{\sqrt{3}}{2} X_0^\prime \, , \nonumber \\
{\tilde X}_0^\prime & \equiv & \left| 1_\qq , 1_\QQ \right>_0 =
+ \frac{\sqrt{3}}{2} X_0 - \frac{1}{2} X_0^\prime \, , \nonumber \\
{\tilde Z} & \equiv & \left| 1_\qq , 0_\QQ \right>_1 =
\frac{1}{\sqrt{2}} \left( Z^\prime \! + Z \right) \, , \nonumber \\
{\tilde Z}^\prime & \equiv & \left| 0_\qq , 1_\QQ \right>_1 =
\frac{1}{\sqrt{2}} \left( Z^\prime \! - Z \right) \, .
\label{eq:HQbasis}
\end{eqnarray}
Expressing the basis change between Eqs.~(\ref{eq:Swavediquark}) and
(\ref{eq:HQbasis}) in terms of rotation matrices, one finds
\begin{eqnarray}
J^{PC} = 0^{++}: & \; & \left( \! \begin{array}{c} X_0 \\ X^\prime_0
\end{array} \! \right) = \left( \begin{array}{cc} \cos \frac{\pi}{3}
& \hphantom{+} \sin \frac{\pi}{3} \\[2pt] \sin\frac{\pi}{3} & -\cos
\frac{\pi}{3} \end{array} \right) \left( \! \begin{array}{cc}
{\tilde X}_0 \\ {\tilde X}^\prime_0 \end{array} \! \right) \, ,
\nonumber \\
J^{PC} = 1^{++}: & & \hspace{1.7em} X_1 = \left| 1_\qq , 1_\QQ
\right>_1 \, , \nonumber \\
J^{PC} = 1^{+-}: & & \, \left( \! \begin{array}{c} Z \\ Z^\prime
\end{array} \! \right) = \left( \begin{array}{cc} \cos \frac{\pi}{4}
& -\sin \frac{\pi}{4} \\[2pt] \sin \frac{\pi}{4} &\hphantom{+}  \cos
\frac{\pi}{4} \end{array} \right) \left( \! \begin{array}{c}
{\tilde Z} \\ {\tilde Z}^\prime \end{array} \right) \, , \nonumber \\
J^{PC} = 2^{++}: & & \hspace{1.7em} X_2 = \left| 1_\qq , 1_\QQ
\right>_2 \, ,
\label{eq:SwaveQQ}
\end{eqnarray}
where outer subscripts again indicate total quark spin $J$.

This nomenclature (adapted from Ref.~\cite{Maiani:2014aja}) applies to
both $I \! = \!  0$ and $I \! = \! 1$ states.  However, its use may
cause confusion because the label $Z$ is usually understood to mean
only $I \! = \! 1$ states, whereas we use $Z$ to mean $1^{+-}$ states
exclusively.  The naming scheme adopted by the Particle Data
Group~\cite{Tanabashi:2018oca} labels the $I \! = \!  0$ $J^{++}$
states as $\chi^{\vphantom\dagger}_J$ and the $I \! = \! 0$ $1^{+-}$
state as $h$, exactly as for conventional quarkonium, while the label
$Z$ is reserved for $I \! = \! 1$ $1^{+-}$ states, and the
yet-unobserved $I \! = \! 1$ $J^{++}$ states are called $W_J$.

The mass eigenstates formed from the states of degenerate $J^{PC}$ in
Eq.~(\ref{eq:Swavediquark}) are defined as
\begin{eqnarray}
\left( \! \begin{array}{c} {\bar X}_0 \\ {\bar X}^\prime_0 \end{array}
\! \right) & = & \left( \! \begin{array}{cc} \hspace{1em} \cos \tX &
\sin \tX \\ -\sin \tX & \cos \tX \end{array}
\! \right) \left( \! \begin{array}{c} X_0 \\ X^\prime_0 \end{array} \!
\right) \, , \nonumber \\
\left( \! \begin{array}{c} {\bar Z} \\ {\bar Z}^\prime \end{array}
\right) & = & \left( \! \begin{array}{cc} \hspace{1em} \cos \tZ &
\sin \tZ \\ -\sin \tZ & \cos \tZ \end{array}
\! \right) \left( \! \begin{array}{c} Z \\ Z^\prime \end{array}
\right) \, .
\label{eq:Mix}
\end{eqnarray}
While it is not logically necessary to require the mixing angles $\tX$
and $\tZ$ for these systems to assume the same values in both the $I
\! = \! 0$ and $I \! = \! 1$ channels, to do so is a reasonable
minimal ansatz.  As shown below, this ansatz does not conflict with
current experimental findings.  In particular, each occurrence of the
label $Z$ below is understood to apply equally well to the $I \!  = \!
0$ or $I \! = \! 1$ state, unless a particular $I$ eigenvalue is
explicitly appended.

The full model Hamiltonian reads
\begin{eqnarray}
H & = & M_0 + 2 \kappa_{qQ} ({\bf s}_q \! \cdot \! {\bf s}_Q +
{\bf s}_{\bar q} \! \cdot \! {\bf s}_{\bar Q}) + V_0 \, {\bm \tau}_q
\! \cdot \! {\bm \tau}_{\bar q} \; {\bm \sigma}_q \! \cdot \!
{\bm \sigma}_{\bar q} \, ,
\label{eq:Ham}
\end{eqnarray}
where $M_0$ is the common multiplet mass, computed in
Ref.~\cite{Giron:2019bcs} using spin- and isospin-blind
Schr\"{o}ding\-er equations that depend only upon the diquark (or
also, in the pentaquark case, triquark) mass and a central potential
computed on the lattice from pure-glue configurations.  The second
term of Eq.~(\ref{eq:Ham}) represents the primary interaction of the
``Type-II'' diquark model~\cite{Maiani:2014aja}, with the parameter
$\kappa_{qQ}$ representing the strength of the spin-spin couplings
within diquarks ($q$ only to $Q$, $\bar q$ only to $\bar Q$).  Note
particularly the assumption that the dominant isospin-dependent
potential in Eq.~(\ref{eq:Ham}) depends only upon the light-quark
spins, rather than the diquark spins; were the diquarks truly
pointlike, then the $q$($\bar q$) would still carry all the isospin of
$\de$($\bde$), but the $V_0$ interaction would be replaced with
\begin{equation}
\Delta H = V_1 \, {\bm \tau}_q \! \cdot \! {\bm \tau}_{\bar q} \;
{\bm \sigma}_\de \! \cdot \! {\bm \sigma}_\bde \, .
\label{eq:IsospinDiquark}
\end{equation}

The particular form of the Hamiltonian Eq.~(\ref{eq:Ham}) [or
(\ref{eq:IsospinDiquark})] deserves further comment.  The original
spin-dependent diquark model for heavy-quark exotics (called ``Type
I''~\cite{Maiani:2004vq}), which emerged when only a few such exotics
were known, allows for couplings between all four quarks.  As exotics
data improved with time, it became apparent that the ``Type-I'' model
was unsuited to describing the full observed spectrum.  The ansatz of
the ``Type-II'' model~\cite{Maiani:2014aja} takes the dominant spin
couplings to be just those within each diquark (since the diquarks are
believed to be more compact than the full hadron), and the model
provides a satisfactory understanding of the masses of the states
$X(3872)$, $Z_c(3900)$, and $Z_c(4020)$---but not, by construction,
their isospins.  The ``Type-II'' ansatz is simplest to justify if one
appeals to the kinematically induced separation of the $\de$-$\bde$
pair in the dynamical diquark picture~\cite{Brodsky:2014xia}, but one
could just as easily suppose the existence of a potential barrier
separating the $\de$-$\bde$ pair~\cite{Maiani:2019cwl}.  A comparison
of the spectra of the ``Type-II'' and molecular models forms the basis
of a dedicated study in Ref.~\cite{Cleven:2015era}.  The $V_0$ term of
Eq.~(\ref{eq:Ham}) or $V_1$ term of Eq.~(\ref{eq:IsospinDiquark})
introduces isospin dependence into the model, and under the assumption
of chiral-type couplings, the isospin exchange is linked to additional
spin dependence.  Certainly one could also incorporate, for example,
isospin-dependent, spin-independent Hamiltonian operators, but for
this initial study the $V_0$ and $V_1$ operators are designed to
assume the most familiar form from chiral dynamics.

The matrix elements of the symmetry-breaking operators in
Eq.~(\ref{eq:Ham}) are computed easily using standard
square-completion tricks.  The second term evaluates to
\begin{equation}
\kappa_{qQ} \left[ s_\de (s_\de + 1) + s_{\bde} (s_{\bde} + 1) - 3
\right] \, ,
\end{equation}
which is trivially computed for states expressed in the diquark basis
of Eq.~(\ref{eq:Swavediquark}), for which the operator is diagonal.
The third term of Eq.~(\ref{eq:Ham}) evaluates to
\begin{equation}
V_0 \left[ 2I(I+1) - 3 \right] \left[ 2s_{\qq}(s_{\qq} +1) - 3 \right]
\, ,
\end{equation}
which is trivially computed for states expressed in the total
light-quark spin ($s_{\qq}$) basis of Eq.~(\ref{eq:HQbasis}).  In the
alternative form of Eq.~(\ref{eq:IsospinDiquark}), one obtains instead
the contribution
\begin{equation} \label{eq:V1}
2V_1\left[2I\left(I+1\right)-3\right] \left[J\left(J+1\right)
-s_{\de}\left(s_{\de}+1\right)-s_{\bde}\left(s_{\bde}+1\right)\right]
\, ,
\end{equation}
which again is easily computed in the diquark-spin basis of
Eq.~(\ref{eq:Swavediquark}).

Using the mass eigenstates defined in Eq.~(\ref{eq:Mix}) and the
Hamiltonian of Eq.~(\ref{eq:Ham}), one immediately computes the masses
for the 12 physical states in the $\Sigma^+_g(1S)$ multiplet:

%\begin{widetext}
\begin{eqnarray}
M^{I=0}_{{\bar X}_0} & = & M_0 - \kappa_{qQ} \left[ 1 + 2 \cos
(2\tX) \right] + 3V_0 \left[ 1 - 2\cos \left(2\tX +
\frac{\pi}{3} \right) \right] \, , \nonumber \\
M^{I=0}_{{\bar X}^\prime_0} & = & M_0 - \kappa_{qQ} \left[ 1 - 2 \cos
(2\tX) \right] + 3V_0 \left[ 1 + 2\cos \left(2\tX +
\frac{\pi}{3} \right) \right] \, , \nonumber \\
M^{I=1}_{{\bar X}_0} & = & M_0 - \kappa_{qQ} \left[ 1 + 2 \cos
(2\tX) \right] - V_0 \left[ 1 - 2\cos \left(2\tX +
\frac{\pi}{3} \right) \right] \, , \nonumber \\
M^{I=1}_{{\bar X}^\prime_0} & = & M_0 - \kappa_{qQ} \left[ 1 - 2 \cos
(2\tX) \right] - V_0 \left[ 1 + 2\cos \left(2\tX +
\frac{\pi}{3} \right) \right] \, , \nonumber \\
M^{I=0}_{X_1} & = & M_0 - \kappa_{qQ} - 3V_0 \, , \nonumber \\
M^{I=1}_{X_1} & = & M_0 - \kappa_{qQ} + V_0 \, , \nonumber \\
M^{I=0}_{X_2} & = & M_0 + \kappa_{qQ} - 3V_0 \, , \nonumber \\
M^{I=1}_{X_2} & = & M_0 + \kappa_{qQ} + V_0 \, , \nonumber \\
M^{I=0}_{\bar Z} & = & M_0 - \kappa_{qQ} \cos (2\tZ) + 3V_0
\left[ 1 - 2\sin (2\tZ) \right] \, , \nonumber \\
M^{I=0}_{{\bar Z}^\prime} & = & M_0 + \kappa_{qQ} \cos (2\tZ) +
3V_0 \left[ 1 + 2\sin (2\tZ) \right] \, , \nonumber \\
M^{I=1}_{\bar Z} & = & M_0 - \kappa_{qQ} \cos (2\tZ) - V_0
\left[ 1 - 2\sin (2\tZ) \right] \, , \nonumber \\
M^{I=1}_{{\bar Z}^\prime} & = & M_0 + \kappa_{qQ} \cos (2\tZ) -
V_0 \left[ 1 + 2\sin (2\tZ) \right] \, .
\label{eq:Master}
\end{eqnarray}
%
%\end{widetext}

These 12 masses depend upon a common multiplet mass $M_0$, two
Hamiltonian parameters ($\kappa_{qQ}$ and $V_0$), and the mixing
angles $\theta_{\! X,Z}$.  At this point, Eqs.~(\ref{eq:Master}) are
equally valid for $c\bar c$ and $b\bar b$ tetraquarks, as well as
$B_c$ tetraquarks if one includes distinct $\kappa_{qc}$ and
$\kappa_{qb}$ couplings.  Note that the primed and unprimed states
interchange under a trivial shift of the mixing angles: ${\bar X}_0 \!
\leftrightarrow \! {\bar X}^\prime_0$ when $\tX \! \to \! \tX \! + \!
\frac{\pi}{2}$, and similarly ${\bar Z} \! \leftrightarrow \!
{\bar Z}^\prime$ when $\tZ \! \to \! \tZ \! + \! \frac{\pi}{2}$.
Therefore, the unprimed and primed states are equally valid for
purposes of parametric fitting to the mass spectrum.  However, these
states remain inequivalent in terms of their content according to
$s_{\qq}$ and $s_{\QQ}$ eigenvalues, a distinction that can be probed
through their decay modes.

If Eq.~(\ref{eq:IsospinDiquark}) is used instead, then the $V_0$ terms
of Eqs.~(\ref{eq:Master}) are replaced by the $V_1$ terms
\begin{eqnarray}
\Delta M^{I=0}_{\bar{X}_0} & = &
+12V_1 \left[1-\cos\left(2\tX \right)\right] \, , \nonumber \\
\Delta M^{I=0}_{\bar{X}'_0} & = &
+12V_1 \left[1+\cos\left(2\tX \right)\right] \, , \nonumber \\
\Delta M^{I=1}_{\bar{X}_0} & = &
-4V_1 \left[1-2\cos\left(2\tX\right)\right] \, , \nonumber \\
\Delta M^{I=1}_{\bar{X}'_0} & = &
-4V_1\left[1+2\cos\left(2\tX\right)\right] \, , \nonumber \\
\Delta M^{I=0}_{X_1} & = & +0 \, , \nonumber \\
\Delta M^{I=1}_{X_1} & = & +0 \, , \nonumber \\
\Delta M^{I=0}_{X_2} & = & -12V_1 \, , \nonumber \\
\Delta M^{I=1}_{X_2} & = & +4V_1 \, , \nonumber \\
\Delta M^{I=0}_{\bar{Z}} & = &
+6V_1\left[1-\cos\left(2\tZ \right)\right] \, , \nonumber \\
\Delta M^{I=0}_{\bar{Z}'} & = &
+6V_1\left[1+\cos\left(2\tZ \right)\right] \, , \nonumber \\
\Delta M^{I=1}_{\bar{Z}} & = &
-2V_1\left[1-\cos\left(2\tZ \right)\right] \, , \nonumber \\
\Delta M^{I=1}_{\bar{Z}'} & = &
-2V_1\left[1+\cos\left(2\tZ \right)\right] \, .
\label{eq:Master2}
\end{eqnarray}
The invariance of ${\bar X}_0 \! \leftrightarrow \! {\bar X}^\prime_0$
under $\tX \! \to \! \tX \! + \! \frac{\pi}{2}$, and ${\bar Z} \!
\leftrightarrow \! {\bar Z}^\prime$ under $\tZ \! \to \! \tZ \! + \!
\frac{\pi}{2}$ also holds in this case.  However, the most conspicuous
feature of Eqs.~(\ref{eq:Master2}) is the degeneracy of $X^{I=0}_1$
and $X^{I=1}_1$.  As these states represent the $X(3872)$ candidate
and its yet-unseen charged partner~\cite{Choi:2011fc}, one finds that
a model in which the diquarks exchange isospin only in their pointlike
form as in Eq.~(\ref{eq:IsospinDiquark}) runs afoul of known
phenomenology.  That these two $V_1$ contributions are not just equal
but indeed zero follows immediately from Eq.~(\ref{eq:V1}) and the
fact [Eq.~(\ref{eq:Swavediquark})] that the $X_1$ states ($J \! = \!
1$) contain only components in which $s_\de \! = \! 0$ and $s_\bde \!
= \!  1$, or vice versa.  We therefore analyze as our minimal model
the Hamiltonian given by Eq.~(\ref{eq:Ham}), which leads to the
spectrum given by Eq.~(\ref{eq:Master}). In addition, if one neglects
all isospin-independent couplings ($\kappa_{qQ} \! = \! 0$), then
Eq.~(\ref{eq:Master}) shows that the isoscalar, spin-2 state
$X_2^{I=0}$ would be degenerate with the $X(3872)$, again in
opposition to known phenomenology.

\section{Results and Analysis}
\label{sec:Results}

We now test whether this model can accommodate what is known about the
ground-state [$\Sigma^+_g(1S)$] hidden-charm exotics, the $J^{PC} \! =
\! 1^{++}$ $X(3872)$, and the $1^{+-}$ states
$Z_c(3900)$ and $Z_c(4020)$ (the $C$ parity eigenvalue referring to
the neutral states).  The Particle Data Group~\cite{Tanabashi:2018oca}
averages for their masses are
\begin{eqnarray}
m_{X(3872)}   & = & 3871.69 \pm 0.17 \ {\rm MeV} \, , \nonumber \\
m_{Z_c(3900)} & = & 3887.2 \pm 2.3   \ {\rm MeV} \, , \nonumber \\
m_{Z_c(4020)} & = & 4024.1 \pm 1.9   \ {\rm MeV}  \, .
\label{eq:Masses}
\end{eqnarray}
The $Z_c$ states have been observed in both charged and neutral
variants, which decay in their closed-charm modes to $\pi^\pm$ and
$\pi^0$, respectively, meaning that they have $I \! = \! 1$ [and hence
$G \! = \!  C (-1)^I \! = \!  +$].  The dominant decays of these
states have open charm: $Z_c(3900) \! \to \! {\bar D}^* D$ and
$Z_c(4020) \! \to \! {\bar D}^* D^*$~\cite{Tanabashi:2018oca}, making
an analysis based solely upon isospin not as incisive.  On the other
hand, we noted above that no charged partner of the $X(3872)$ has been
observed despite a dedicated search~\cite{Choi:2011fc}, which suggests
$I \! = \! 0$.  However, the $X(3872)$ is widely believed to be unique
among all known hadrons in possessing a valence quark content (more
$c\bar c u\bar u$ than $c\bar c d\bar d$) not corresponding to just
one $I$ eigenvalue.  Its mass is almost precisely equal that of ${\bar
D}^{*0} D^0$ (in fact, its dominant decay mode is ${\bar D}^{*0}
D^0$~\cite{Tanabashi:2018oca}) but about 8~MeV below that of ${\bar
D}^{*+} D^-$, while the idealized $I \! = \! 0$ and $I \! = \!  1$
combinations of these states are equal admixtures.  Likewise,
$X(3872)$ has been observed to decay to both the $G \! = \! -$ (hence
$I \!  = \! 1$) final state $\pi^+ \pi^- J/\psi$ and the $G \! = \! +$
final states $\omega J/\psi$ ($I \! = \! 0$) and (very
recently~\cite{Ablikim:2019soz}) $\pi^0 \chi_{c0}$ ($G \! = \! -$, $I
\! = \! 1$).  The mere facts that $m_u \!  < \! m_d$ (and $q_u \!
\neq q_d$) and that $X(3872)$ lies in the close proximity to the
threshold for one particular charge combination appear to be
responsible for these fascinating results.  In the current model,
however, we take $m_u \! = \! m_d$, ignore electromagnetic effects,
and treat $X(3872)$ as the unique $I \! = \!  0$ $1^{++}$ state in
$\Sigma^+_g(1S)$, $X_1^{I=0}$.  Indeed, the same analysis below holds
even if the $X(3872)$ is an exact $I \! = \! 0$ state (an equal
admixture of $c\bar c u\bar u$ and $c\bar c d\bar d$), and its
isospin-violating decays are purely the result of the kinematical
blocking of the $d\bar d$ channel.

Without performing a detailed accounting of every significant source
of fine-structure splitting expected to appear in these states, a
precise estimate of the numerical uncertainties on our mass
predictions is impossible.  Nevertheless, if the model is to have any
validity, it must incorporate basic phenomenological facts such as the
$\sim \! 20$~MeV mass difference $m_{Z_c(3900)} \! - \!  m_{X(3872)}$.
One may therefore take 20~MeV as a reasonable upper limit for mass
uncertainties in this model.

Using the values in Eqs.~(\ref{eq:Masses}) (with uncertainties
suppressed) in Eqs.~(\ref{eq:Master}), one obtains
%
%\begin{widetext}
\begin{eqnarray}
\frac 1 2 \left( m_{Z_c(4020)} + m_{Z_c(3900)} \right) & = & M_0 - V_0
= 3955.65 \ {\rm MeV} \, , \nonumber \\
\frac 1 2 \left( m_{Z_c(4020)} + m_{Z_c(3900)} \right) - m_{X(3872)}
& = & \kappa_{qc} + 2V_0 = 83.96 \ {\rm MeV} \, , \nonumber \\
\frac 1 2 \left( m_{Z_c(4020)} - m_{Z_c(3900)} \right) & = &
\left| \kappa_{qc} \cos 2\tZ - 2V_0 \sin 2\tZ \right|
= 68.45 \ {\rm MeV} \, .
\label{eq:Nums}
\end{eqnarray}
%\end{widetext}
%
The absolute value in the third expression reflects the fact, noted
above, that $Z_c(3900)$ and $Z_c(4020)$ may be identified with $\bar
Z_{I \! = \! 1}$ or ${\bar Z}^\prime_{I \! = \! 1}$ in either order,
under the replacement $\tZ \! \to \! \tZ \! + \! \frac{\pi}{2}$.
According to Eqs.~(\ref{eq:HQbasis})--(\ref{eq:Mix}), this
substitution exchanges the relative amounts of the $s_{\QQ} \! = \! 0$
and $s_{\QQ} \! = \!  1$ components in the mass eigenstates.  In
particular, $\tZ \! = \! \frac{\pi}{4}$ takes $\bar Z$ to the pure
$s_{\QQ} \! = \! 0$ eigenstate $\tilde Z$ and takes ${\bar Z}^\prime$
to the pure $s_{\QQ} \! = \! 1$ eigenstate ${\tilde Z}^\prime$, while
$\tZ \! = \! \frac{3\pi}{4}$ takes $\bar Z \! \to \! {\tilde
Z}^\prime$ and ${\bar Z}^\prime \! \to \! \tilde Z$.  Since
$Z_c(3900)$ has been observed to decay to $\pi J/\psi$ and not $\pi
h_c$, while the reverse is true for
$Z_c(4020)$~\cite{Tanabashi:2018oca}, identifying $\bar Z_{I \! = \!
1}$ with $Z_c(3900)$, and ${\bar Z}^\prime_{I \! = \! 1}$ with
$Z_c(4020)$, is best achieved through values $\tZ \! \approx \!
\frac{3\pi}{4}$.  Even so, we do not impose this constraint on the
fit, focusing initially only upon the mass spectrum.

First, even without information on $\tZ$, we predict
\begin{equation}
M^{I=1}_{X_2} = m_{Z_c(4020)} + m_{Z_c(3900)} - m_{X(3872)} = 4039.61
\  {\rm MeV} .
\end{equation}
Should a charged, $J^{PC} \! = \! 2^{++}$ exotic state fail to occur
in the vicinity of 4040~MeV, then the validity of this simplified
model must be reassessed.  Note that the $X_2^{I=1}$ has $G \! =
\! -1$ and, according to Eq.~(\ref{eq:SwaveQQ}), preferentially
decays to $J/\psi$, which also carries $G \! = \! -1$.  This state
would therefore most easily be seen in the channel $\pi \pi J/\psi$.

Imposing the constraints of Eqs.~(\ref{eq:Nums}) on the last 8 mass
expressions in Eq.~(\ref{eq:Master}) leaves the remaining four
non-scalar states, $M^{I=1}_{X_1}$, $M^{I=0}_{X_2}$, $M^{I=0}_{\bar
Z}$, and $M^{I=0}_{{\bar Z}^\prime}$, as functions of the single
parameter $\tZ$.  Taking $\tZ \! \to \! \tZ \! + \! \pi$ simply
changes the normalization sign of both $\bar Z$ and ${\bar Z}^\prime$,
so one may consider only the range $\tZ \! \in \! [0,\pi]$.  The most
important constraint from a phenomenological perspective is that the
``charged partner'' to the $X(3872)$, $X^{I=1}_1$, must be
substantially heavier, at least 20~MeV~\cite{Choi:2011fc}, than the
$X(3872)$.  From Eqs.~(\ref{eq:Master}), one notes that this
constraint simply reads $V_0 \! > \! 5$~MeV\@.  While $V_0$ is not yet
fixed at this stage of the fit, one notes that
\begin{equation}
\frac 1 2 \left( M^{I=1}_{X_1} + M^{I=0}_{X_2} \right) = M_0 - V_0 =
3955.65 \ {\rm MeV} \, , 
\label{eq:HardAvg}
\end{equation}
meaning that allowing $X^{I=1}_1$ to be excessively heavy forces the
spin-2 isoscalar $X^{I=0}_2$ to be so light that it would already have
been observed.  $\tZ$ can only be allowed in certain numerical ranges
to avoid this problem, but fortunately, these ranges are substantial:
The lighter of $\{ X^{I=1}_1$, $X^{I=0}_2 \}$ exceeds the $X(3872)$
mass for $\tZ / \pi \! \in \! [0, 0.10]$, $[0.65, 0.85]$, and $[0.90,
1.00]$.  $X^{I=1}_1$ is the heavier of these two states in the middle
interval and the lighter in the other two intervals.  Within these
ranges, $M^{I=1}_{X_1} \! - m_{X(3872)} \! > \! 20$~MeV for the
restricted ranges $\tZ / \pi \! \in \! [0, 0.04]$, $[0.65, 0.85]$, and
$[0.91, 1.00]$.  The masses of the two isoscalar partners $\{ {\bar
Z}^{\vphantom\prime}_{I=0}, {\bar Z}^\prime_{I=0} \}$ to the
$Z_c(3900)$ and $Z_c(4020)$ exceed $m_{X(3872)}$ over these full
ranges, and the lighter of the two exceeds $m_{Z_c(3900)}$ over the
restricted ranges except for the small interval $\tZ/\pi \! \in \!
[0.71, 0.75]$, where even there it is never more than about 2~MeV
below $m_{Z_c(3900)}$; indeed, precisely at the ideal mixing angle
$\tZ \! = \! \frac{3\pi}{4}$, Eqs.~(\ref{eq:Master}) show that the
isoscalar ${\bar Z}^\prime_{I=0}$ becomes degenerate with ${\bar
Z}_{I=1} \!  = \!  Z_c(3900)$.  One finds, therefore, that rather
large ranges of $\tZ$ appear to all satisfy spectroscopic constraints.

The possibility of an isoscalar $1^{+-}$ state quasi-degenerate with
the $Z_c(3900)$ is interesting in light of phenomenological mystery:
$m_{Z_c(3900)}$ as determined via its $\pi J/\psi$ decay channel (pure
$I \! = \! 1$) tends to lie several MeV above its value as determined
through $(D {\bar D}^*)^0$ (a mixture of $I \! = \! 0,
1$)~\cite{Tanabashi:2018oca}.  If the latter resonance turns out to be
a mixture of ${\bar Z}^\prime_{I=0}$ and ${\bar Z}_{I=1}$, then a
shifted mass---an average of the two mass eigenvalues---might be
expected.  In addition, if ${\bar Z}_{I=1}$ is nearly ideally mixed to
decay to $s_{c\bar c} \! = \! 1$ charmonium, then ${\bar
Z}^\prime_{I=0}$ is nearly ideally mixed to decay to $s_{c\bar c} \! =
\! 0$ charmonium, meaning that one would have a mixture of both
components in this scenario.  However, since the $Z_c(3900)^\pm$ mass
measured through the channel $(D {\bar D}^*)^\pm$ (a pure $I \! = \!
1$ combination) is also low, this resolution is not entirely
satisfactory.  Nevertheless, such mixing should be kept in mind as a
possibility, should the $Z_c(3900)$ mass discrepancy persist.

Turning to the decay properties, we have already noted the
preferential coupling of $Z_c(3900)$ to $J/\psi$ and $Z_c(4020)$ to
$h_c$.  As easily seen from combining
Eqs.~(\ref{eq:HQbasis})--(\ref{eq:Mix}), the $s_{c\bar c} \! = \! 1$
content of ${\bar Z}_{I=1} \! = \!  Z_c(3900)$ is given by
\begin{equation} \label{eq:Zcontent}
P_{s_{c\bar c} = 1} [Z_c(3900)] = \sin^2 \! \left( \tZ - \frac{\pi}{4}
\right) \, .
\end{equation}
In the restricted allowed ranges for $\tZ / \pi$ listed above, we find
$P_{s_{c\bar c} = 1} [Z_c(3900)] \! \in \! [0.36, 0.50]$, $[0.91,
1.00]$, and $[0.50, 0.76]$, respectively.  In light of the preference
for $Z_c(3900) \! \to \! J/\psi$ and $Z_c(4020) \! \to
\!  h_c$, the second region, $\tZ/\pi \! \in \! [0.65, 0.85]$,
appears to be favored.  Note a very recent
result~\cite{Ablikim:2019ipd}, the observation of $Z_c(3900)^\pm \!
\to \! \rho^\pm \eta_c$, indicating that $Z_c(3900)$ is not a perfect
$s_{c\bar c} \! = \! 1$ state.

We now consider the scalar sector.  Recalling that ${\bar X}_0 \!
\leftrightarrow \! {\bar X}^\prime_0$ when $\tX \! \to \! \tX \! + \!
\frac{\pi}{2}$ for both the $I \! = \! 0, 1$ channels, one need
consider only the range $\tX/\pi \! \in \!  [0, \frac 1 2 ]$.  One
then finds, over the preferred range $\tZ/\pi \! \in \! [0.65, 0.85]$
that all four of the ${\bar X}_0$ states are heavier than the
$X(3872)$ over the full range of $\tX$ except in the interval $\tX/\pi
\! \in [0.29, 0.42]$, and in that range only the state ${\bar
X}_0^{\prime I=0}$ is too light.  The other scalar states tend to be
much heavier, ranging from at least 3900~MeV to well over 4200~MeV\@.
In summary, mixing angles in the ranges
\begin{equation}
\frac{\tX}{\pi} \in [0, 0.29], \ [0.42, 0.79], \ [0.92, 1] \, , \
\frac{\tZ}{\pi} \in [0.65, 0.85] \, ,
\end{equation}
appear to produce no conflicts with experiment.

In order to demonstrate the full predictive power of the model, we now
choose one allowed set of $\{ \tX, \tZ \}$ and present the complete
set of mass eigenvalues for all 12 states in the ground-state
$\Sigma^+_g(1S)$ multiplet.  We fix $\tZ/\pi \! = \! 0.80$, in which
case [by Eq.~(\ref{eq:Zcontent})] the $Z_c(3900)$ is over 97\%
$s_{c\bar c} \!  = \! 1$, and the original model parameters of
Eq.~(\ref{eq:Ham}) are determined separately as
\begin{equation} \label{eq:FinalParams}
M_0 = 3988.75 \, {\rm MeV} , \ \kappa_{qc} = 17.76 \, {\rm MeV} , \
V_0 = 33.10 \, {\rm MeV} .
\end{equation}
Since $\left< e^{-m r} /r \right> \! \approx \! 3.1$~fm$^{-1}$ for the
$\Sigma^+_g(1S)$ states using $m \! = \! m_\pi$ (while for $m \!  = \!
0.5$~GeV, the value is $\approx \! 2.0$~fm$^{-1}$),
Eq.~(\ref{eq:V0tilde}) gives
\begin{equation}
{\tilde V}_0 = 11.0 \ {\rm MeV \cdot fm} \, ,
\end{equation}
comparable in magnitude to, but a factor 2.3 larger than, the
color-singlet $NN\pi$ coupling of Eq.~(\ref{eq:PionCoeff}).  The
numerical value obtained for $\kappa_{qc}$ in
Eq.~(\ref{eq:FinalParams}) is also interesting in light of the much
larger corresponding value 67~MeV obtained from a fit to exotics in
the (isospin-independent) ``Type-II'' model~\cite{Maiani:2014aja}.
Clearly, much of the strength of the coupling $\kappa_{cq}$ in the
current fit has migrated to the coupling $V_0$.  Indeed,
Ref.~\cite{Maiani:2014aja} notes that a fit to $\kappa_{qc}$ from the
$\Sigma_c$-$\Lambda_c$ mass difference~\cite{Maiani:2004vq} gives
$\kappa_{qc} \! = \! 22$~MeV, which agrees much better with the result
in Eq.~(\ref{eq:FinalParams}).  Additionally setting $\tX/\pi \! = \!
0.49$ to fix the scalar sector, we obtain the full results presented
in Table~\ref{tab:Nums}.  If $M_0$ in Eq.~(\ref{eq:FinalParams}) is
used instead of $m_{X(3872)}$ for the $\Sigma^+_g(1S)$ mass eigenvalue
in a fit such as in Ref.~\cite{Giron:2019bcs} or the right-hand
columns of Table~\ref{tab:FiniteDiquark}, one obtains a diquark mass
$m_\de \!  = \! 1.92$--$1.94$~GeV, about 3\% larger, while the length
scales $\left< 1 / r \right>^{-1}$ or $\left< r \right>$ are about
1.5\% smaller.

\begin{table}
\caption{Masses (in MeV) of the 12 ground-state multiplet
[$\Sigma^+_g(1S)$] states in the dynamical diquark model for the
choice of mixing parameters $\tX \! = \! 0.49 \, \pi$ and $\tZ \! = \!
0.80 \, \pi$.  Experimental inputs [Eqs.~(\ref{eq:Masses})] are in
boldface.}
\vskip 1ex
\centering
\setlength{\extrarowheight}{0.8ex}
\setlength{\tabcolsep}{1em}
\begin{tabular}{ccc}
\hline
$M^{I=0}_{{\bar X}_0} = 4215.7$ &
$M^{I=0}_{X_1} = \bm{3871.7}$ &
$M_{\bar Z}^{I=0} = 4271.5$ \\
$M^{I=0}_{{\bar X}^\prime_0} = 3924.9$ &
$M^{I=1}_{X_1} = 4004.1$ &
$M_{{\bar Z}^\prime}^{I=0} = 3904.7$ \\
$M^{I=1}_{{\bar X}_0} = 3936.7$ &
$M^{I=0}_{X_2} = 3907.2$ &
$M_{\bar Z}^{I=1} = \bm{3887.2}$ \\
$M^{I=1}_{{\bar X}^\prime_0} = 3939.1$ &
$M^{I=1}_{X_2} = 4039.6$ &
$M_{{\bar Z}^\prime}^{I=1} = \bm{4024.1}$ \\
\hline
\end{tabular}
\label{tab:Nums}
\end{table}

As promised, the lowest state in this multiplet is the $X(3872)$.  Its
``charged partner'' $X_1^{I=1}$ lies a full 130~MeV higher in mass,
and therefore would be expected to be quite wide, possibly
unobservably so.  The price for achieving this gap was noted in
Eq.~(\ref{eq:HardAvg}), that the $X_2^{I=0}$ mass must be pushed
lower, in our example to 3907.2~MeV\@.  In fact, the $\chi_{c2}(3930)$
has the same quantum numbers, and while expected to be the
conventional charmonium $\chi_{c2}(2P)$ state, its most recent mass
measurement by LHCb~\cite{Aaij:2019evc} of $3921.9 \! \pm \! 0.6 \!
\pm 0.2$~MeV is rather lower than earlier
determinations~\cite{Tanabashi:2018oca}, possibly pointing to a more
complicated configuration such as two peaks, or a mixture of
$\chi_{c2}(2P)$ with a tetraquark state.  The possible
quasi-degeneracy of ${\bar Z}^\prime_{I=0}$ with $Z_c(3900)$ has been
noted above.  The $2^{++}$ state $X_2^{I=1}$ lies near the unconfirmed
$C \!= \! +$ state $Z_c(4055)^\pm$, as well as the unconfirmed
charged~\cite{Ablikim:2017oaf} and neutral~\cite{Ablikim:2017aji}
``charmoniumlike structures'' around 4035~MeV\@.

In this particular fit, the scalar mixing angle $\tX$ was chosen to
make the $0^{++}$ state ${\bar X}^{\prime \, I=0}_0$ light, so as to
identify it with the $\chi_{c0}(3915)$.  The nature of this state
remains quite controversial~\cite{Olsen:2019lcx}; for instance, it
might even be the lowest $c\bar c s\bar s$ state~\cite{Lebed:2016yvr}.
Indeed, a very recent determination of the mass of this state as an
$\omega J/\psi$ resonance~\cite{Ablikim:2019zio} gives $3926.4
\! \pm \! 2.2$~MeV\@.  Meanwhile, the states ${\bar X}^{(\prime) \,
I=1}_0$ are quasi-degenerate, appearing near the unconfirmed state
$X(3940)$.  The candidate states above 4200~MeV are very possibly too
wide to resolve experimentally.  Other choices of $\tX$ can push up
all of the scalar states to at least 3950~MeV, or seek to accommodate
the $Z_c^\pm (4100)$ or $X(4160)$, neither of which has been
confirmed, let alone confirmed to have positive parity.  The only
other positive-parity states in this range, $Y(4140)$ and $Y(4274)$,
are ignored in this analysis since they have only been observed as
$\phi J/\psi$ resonances and therefore are very possibly $c\bar c
s\bar s$~\cite{Lebed:2016yvr,Giron:2019bcs}.

The dominant decay modes for exotics of these $J^{PC}$ quantum
numbers~\cite{Cleven:2015era} are expected to be the $S$-wave
open-charm combinations $D\bar D$ and $D^* {\bar D}^*$ for $0^{++}$,
$D {\bar D}^*$ for $1^{++}$, $D {\bar D}^*$ and $D^* {\bar D}^*$ for
$1^{+-}$, and $D^* {\bar D}^*$ for $2^{++}$ (plus charge-conjugate
modes, and for both $I\! = \! 0$ and $I \! = \! 1$).  Since $2m_D \!
\approx \!  3860$~MeV, $m_D \! + \! m_{D^*} \! \approx 3875$~MeV, and
$2m_{D^*} \! \approx \! 4020$~MeV, most of these channels are open for
the corresponding states in Table~\ref{tab:Nums}.  The dominant
closed-charm modes are easily determined from Eqs.~(\ref{eq:HQbasis})
and (\ref{eq:SwaveQQ}) plus $G$-parity conservation.  For example,
$X_2^{I \! = \! 0}$ lies below the $D^* {\bar D}^*$ threshold, and
thus decays into $\omega J/\psi$, but the $D$-wave decay into $D\bar
D$ is also possible, just as for the $\chi_{c2}(3930)$.

Lastly, we note from Table~\ref{tab:Nums} that the full fine-structure
splitting of the $\Sigma^+_g(1S)$ multiplet can be much larger than
the crude estimate of 150~MeV given in Ref.~\cite{Giron:2019bcs}.
However, if the states heavier than 4200~MeV turn out to be
unobservably wide, then the spectrum of {\em observable\/}
$\Sigma^+_g(1S)$ states does indeed turn out to be about 170~MeV\@.

\section{Discussion and Conclusions}
\label{sec:Discussion}

In this paper we have developed a variant of the dynamical diquark
model in which isospin dependence is explicitly incorporated.  We also
developed a simple modification of our calculation to test for effects
on the exotics spectrum due to finite diquark size.

Allowing for the potential to transition from one describing the
interaction of diquarks at large separations $r$ to one dominated by
the $\QQ$ interaction at separations below a chosen value $r \! = \!
R$ effectively introduces an effective diquark radius of $R/2$.  We
find by explicit calculation that the exotics spectrum changes very
little until $R$ is as large as 0.8~fm, meaning that results obtained
by treating the diquarks as pointlike are reliable even for compact
diquarks with radii as large as 0.4~fm.

The existence of an isospin-dependent interaction between separated
diquarks, a type of (partly) colored pion exchange, is inspired by the
existence of a Nambu-Goldstone theorem of chiral-symmetry breaking
shown to occur in dense QCD\@.  Isospin dependence is clearly evident
in the observation of exotic states to appear in isosinglets and
isotriplets, rather than quartets.  We applied this ansatz of isospin
dependence to the 12 states in the ground-state multiplet
$\Sigma^+_g(1S)$, taking $X(3872)$, $Z_c(3900)$, and $Z_c(4020)$ as
members, and predicted the masses of the others.

The $X(3872)$ in this model naturally emerges over large portions of
the allowed parameter space as the lightest exotic state, and its
``charged partner'', the $J^{PC} \! = \! 1^{++}$ $I \! = \! 1$ member
of the multiplet, is much heavier.  Moreover, the decay preferences
$Z_c(3900) \! \to \! J/\psi$ and $Z_c(4020) \! \to \! h_c$ emerge
directly from the analysis of the mass spectrum.  We have obtained
fits in which several of the ill-characterized low-lying exotics
naturally appear as members of the $\Sigma^+_g(1S)$ multiplet, and
some of the predicted mass eigenvalues lie so high above the dominant
``fall-apart'' decay mode of the corresponding state that they may be
too wide to discern easily.

The natural next step is to consider the first excited multiplet,
$\Sigma^+_g(1P)$, whose states all carry negative parity.  A number of
states have been assigned to this multiplet~\cite{Giron:2019bcs}, such
as $Y(4220)$ and $Y(4360)$.  However, both experimental and
theoretical issues complicate this analysis.  This mass region
includes the expected location of the lightest hybrid charmonium
states~\cite{Liu:2012ze}, which lie outside this analysis.
Additionally, no $P \! = \! -$, $I \! = \! 1$ exotic states have yet
been confirmed.  From the perspective of modeling, several other
operators not included in Eq.~(\ref{eq:Ham}) need to be considered,
not least of which are the tensor operator $S_{12}$ of
Eq.~(\ref{eq:Tensor}) (both isospin-dependent and independent) and the
spin-orbit operator.  In the second excited multiplet [including
states such as $Z_c(4430)$], one expects the range of masses of states
in either $\Sigma^+_g(2S)$ or $\Sigma^+_g(1D)$ to
overlap~\cite{Giron:2019bcs}, or even for the states themselves in the
two multiplets to mix via tensor terms, again complicating the
analysis.  In short, not enough states have been fully characterized
in the excited multiplets of this model to perform a reliable
analysis.

Nevertheless, one basic feature is expected to hold for the excited
multiplets: With reference to Eq.~(\ref{eq:V0tilde}), excited states
are spatially larger, meaning that $\left< e^{-m_\pi r}/r
\right>$ is smaller in higher levels, and so one expects smaller mass
splittings within the higher multiplets in this model (an analogous
effect occurs for fine-structure splittings in ordinary quarkonium).

Multiplets of exotics with excited glue fields, such as
$\Pi^+_u(1P)$-$\Sigma^-_u(1P)$, have not even been mentioned in this
paper, since as was shown in Ref.~\cite{Giron:2019bcs}, they are
expected to lie about 1~GeV above the $\Sigma^+_g(1S)$ ground states
(just like the gap between quarkonium hybrids and conventional
quarkonium states).  Nevertheless, were they to be considered in a
model analogous to the one described here, yet further operators would
need to be included, such as ones dependent upon not only quark spin,
but the spin of the nontrivial glue degrees of freedom as
well~\cite{Brambilla:2018pyn}.

Lastly, all of the phenomenology presented here refers to the sector
of hidden-charm tetraquarks.  In the pentaquark sector, the states
according to the model of Ref.~\cite{Lebed:2015tna} contain triquarks
of the form $\bt \! = \! [\bar c (ud)_{\bar{\bm 3}}]_{\bm 3}$, where
the $ud$ pair is an $I \! = \! 0$ diquark inherited from the initial
$\Lambda_b$ decay process from which all pentaquarks to date have been
produced.  But in that case, the $\bt$-$\de$ pair does not exchange
isospin, and only simple $I \! = \! \frac 1 2$ pentaquarks occur.
Likewise, if the $ud$ diquark carries spin 0 like that in $\Lambda_b$,
then the triquark uniquely carries spin $\frac 1 2$.  Transitions to
the higher-mass ``bad'' ($I \! = \! 1$, spin-1) light diquark are
certainly possible, but are expected to be suppressed.  One may then
study the pentaquarks in a spin-only formulation of the model, as in,
{\it e.g.}, Ref.~\cite{Zhu:2015bba}, or using a different
diquark-triquark formulation as in Ref.~\cite{Ali:2019clg}.

Nor has the $b$ sector been discussed in this paper.  Again, not
enough states have been observed to attempt a reliable fit to the full
spectrum, but in this case the best-characterized exotic candidates
are isotriplets, the $Z_b(10610)$ and $Z_b(10650)$.  The relative
spacings of $B^{(*)}{\bar B}^{(*)}$ thresholds and conventional
bottomonium levels are different from those in the $c\bar c$ system,
leading to a rather different phenomenology.  The bottom analogue to
the $X(3872)$, the $I \! = \! 0$ $1^{++}$ state $X_b$ (see
Refs.~\cite{Guo:2014sca,Karliner:2014lta} for nice discussions of its
expected properties) has not yet been observed.  While $X(3872)$
emerged naturally as the lightest state among the hidden-charm
tetraquarks in this model, there exist alternate portions of the
parameter space where the $I \! = \! 0$ $1^{++}$ state is not the
lightest, and this observation may turn out to be relevant for the $b$
system.

\begin{acknowledgments}
  R.F.L.\  and J.F.G.\ were supported by the National Science
  Foundation (NSF) under Grant No.\ PHY-1803912; J.F.G.\ also
  received support through the Western Alliance to Expand Student
  Opportunities (WAESO) Louis Stokes Alliance for Minority
  Participation Bridge to the Doctorate (LSAMPBD) NSF Cooperative
  Agreement HRD-1702083; and C.T.P.\ through a NASA traineeship grant
  awarded to the Arizona/NASA Space Grant Consortium.
\end{acknowledgments}

\bibliographystyle{apsrev4-1}
\bibliography{diquark}
\end{document}